\documentclass[10pt,conference]{IEEEtran}
\IEEEoverridecommandlockouts
\usepackage{cite}
\usepackage{amsmath,amssymb,amsfonts}
\usepackage{algorithmic}
\usepackage{graphicx}
\usepackage{textcomp}
\usepackage{xcolor}
\usepackage{braket}
\usepackage{quantikz}
\def\BibTeX{{\rm B\kern-.05em{\sc i\kern-.025em b}\kern-.08em
    T\kern-.1667em\lower.7ex\hbox{E}\kern-.125emX}}
\usepackage[colorlinks=true,urlcolor=blue, citecolor=blue, linkcolor=blue, plainpages=false]{hyperref}  
\makeatletter
\newcommand{\linebreakand}{%
  \end{@IEEEauthorhalign}
  \hfill\mbox{}\par
  \mbox{}\hfill\begin{@IEEEauthorhalign}
}
\makeatother

\begin{document}

\title{Mid-Circuit Measurements for Clifford Noise Reduction in Hamiltonian Simulations \\
}

\author{\IEEEauthorblockN{James Brown}
\IEEEauthorblockA{\textit{qBraid Co.} \\
Chicago, IL, USA \\
james@qbraid.com}
\and
\IEEEauthorblockN{Jason Iaconis}
\IEEEauthorblockA{\textit{IonQ Inc.} \\
College Park, MD, USA \\
iaconis@ionq.co}
\and
\IEEEauthorblockN{Yuri Alexeev}
\IEEEauthorblockA{\textit{NVIDIA Corporation} \\
Santa Clara, CA, USA \\
yalexeev@nvidia.com}
\and
\IEEEauthorblockN{Linta Joseph}
\IEEEauthorblockA{\textit{qBraid Co.} \\
Chicago, IL, USA \\
linta@qbraid.com}
\and
\IEEEauthorblockN{Spencer Churchill}
\IEEEauthorblockA{\textit{IonQ Inc.} \\
College Park, MD, USA \\
churchill@ionq.co}
\linebreakand
\IEEEauthorblockN{Kenny Heitritter}
\IEEEauthorblockA{\textit{qBraid Co.} \\
Chicago, IL, USA \\
kenny@qbraid.com}
\and
\IEEEauthorblockN{William Aguilar-Calvo}
\IEEEauthorblockA{\textit{qBraid Co.} \\
Chicago, IL, USA \\
william.aguilar@qbraid.com}
\and
\IEEEauthorblockN{Martin Roetteler}
\IEEEauthorblockA{\textit{IonQ Inc.} \\
College Park, MD, USA \\
martin.roetteler@ionq.co}
\and
\IEEEauthorblockN{Martin Suchara}
\IEEEauthorblockA{\textit{IonQ Inc.} \\
College Park, MD, USA \\
martin.suchara@ionq.co}
}

\maketitle

\begin{abstract}
Quantum simulation of fermionic Hamiltonians is a leading application of quantum computing, but accurate execution on present-day hardware is limited by error accumulation in deep Trotter circuits. We present a device-matched noise-reduction framework for encoded Hamiltonian simulation that combines symplectic-transvection-based Trotter synthesis in the Generalized Superfast Encoding (GSE) with Clifford Noise Reduction (CliNR) and Shor-style stabilizer verification enabled by mid-circuit measurement. We implement this approach for a six-qubit encoded Clifford Trotter step on a Barium development system similar to the forthcoming IonQ Tempo line and benchmark it against direct execution using both hardware experiments and a calibrated device-level noise model. The encoded CliNR execution achieves up to 54\% lower logical error rate. Crucially, this advantage disappears when stabilizer readout is deferred to the end of the circuit, showing that timely mid-circuit fault detection, rather than verification overhead alone, drives the improvement. As a proof of concept, we further show that machine-learning-guided stabilizer selection can identify verification operators that outperform random choices. These results demonstrate that encoding-native verification combined with dynamic-circuit primitives can materially improve application-motivated quantum simulation without the full overhead of quantum error correction.
\end{abstract}


\section{Introduction}

Quantum simulation and electronic-structure calculation are widely regarded as leading applications of quantum computing because they target many-body problems whose classical cost grows rapidly with system size \cite{lloyd1996universal,kassal2008polynomial,mcArdle2020quantum,daley2022practical}. In principle, quantum hardware can represent and evolve such systems more naturally, opening a route to practically relevant calculations in chemistry and materials. This is especially true for chemically and biologically relevant dynamical problems, including coherent energy transport in systems such as the Fenna--Matthews--Olson complex, which require reliable simulation of fermionic Hamiltonians over many time steps \cite{engel2007evidence,panitchayangkoon2010longlived}. On present-day devices, however, noise accumulates over those long circuits and can substantially distort the simulated dynamics.

In this work, we study how to limit the spread of errors in encoded quantum simulation circuits. Our focus is on the regime now opened by rapidly improving hardware capabilities, where dynamic-circuit primitives such as mid-circuit measurement can be combined with encoding-aware circuit design to suppress errors before they contaminate the simulated dynamics~\cite{campbell2024series,decross2023qubit,hothem2025measuring}. The central challenge is therefore not merely to mitigate noise in the abstract, but to identify the error-detection and noise-reduction strategy best matched to a given device, encoding, and circuit structure.

Several complementary strategies can be used to arrest the spread of noise. Continued improvements in gate fidelity reduce the rate at which faults enter the circuit, while compilation, error-mitigation, and partial verification schemes reduce their effect without the full overhead of fault tolerance~\cite{cai2023quantum}. Mid-circuit measurement further expands this design space by enabling in-circuit syndrome extraction, ancilla reuse, and early rejection of faulty runs~\cite{decross2023qubit,hothem2025measuring}. In the longer term, quantum error correction remains the principled route to scalable accuracy~\cite{campbell2024series}, but its overhead motivates approaches that exploit structure already present in the target Hamiltonian. For fermionic simulation, local encodings such as the Generalized Superfast Encoding (GSE) are especially attractive because they replace nonlocal Jordan--Wigner strings with low-weight operators and stabilizers that better match hardware constraints \cite{setia2019superfast,bravyi2002fermionic,whitfield2011simulation,verstraete2005mapping}.

Within this landscape, Clifford Noise Reduction (CliNR) is particularly well suited to the Clifford subcircuits that arise in Trotterized Hamiltonian simulation \cite{delfosse2024lowcost}. Rather than applying those Clifford operations directly to the data register, CliNR verifies a resource state through stabilizer measurements and then teleports the verified operation onto the computation. This off-state verification is appealing for suppressing hook errors generated by two-qubit gate sequences, since faults can be detected before they spread through the code block. Mid-circuit measurement is central to this mechanism: it enables timely stabilizer checks and ancilla reuse, and it allows faulty resource preparations to be rejected before the subsequent evolution is applied.

GSE provides a natural setting in which to realize this idea \cite{brown2025efficient,setia2019superfast}. Its multi-edge graph construction yields local Majorana operators and loop stabilizers whose weight and overlap properties are well suited to stabilizer-based verification. Because those stabilizers arise directly from the interaction graph, the encoding brings with it a natural error-detecting structure rather than an externally imposed one. In our setting, GSE therefore serves simultaneously as the fermionic mapping and as the verification scaffold that supports CliNR.

To implement stabilizer verification fault-tolerantly, we use Shor-style stabilizer readout, which measures selected stabilizers through ancilla-assisted parity checks designed to limit additional error propagation~\cite{shor1996faulttolerant,tansuwannont2023adaptive}. This choice is important in an encoded setting because naive stabilizer extraction can itself introduce correlated faults. The choice of which stabilizers to measure is then a separate optimization problem: stronger verification improves fault detection, but each additional check increases circuit width and depth. Recent work suggests that machine learning can reduce the cost of error mitigation and related optimization tasks~\cite{liao2024machine}, motivating its use here to predict high-value stabilizer checks from the structure of the compiled Clifford circuit rather than relying solely on brute-force search or unguided sampling.

We validate this framework using both noisy simulation and a Barium development system similar to the forthcoming IonQ Tempo line which has mid-circuit measurement capability. Our simulation stack incorporates a calibrated tempo-1 device-level noise model so that algorithmic behavior can be compared directly with hardware data. This combination of experiment and calibrated simulation allows us to identify both the benefits of the protocol and the hardware effects that are not captured by simplified noise models.

This experimental setting is also well matched to the recent development of CliNR. Prior work established breakeven performance on a 36-qubit trapped-ion processor without mid-circuit measurements \cite{tham2025optimized}, while recent progress across platforms has pushed error-detecting and error-correcting primitives closer to practically useful operating regimes~\cite{campbell2024series,sivak2023realtime}. Here we move from random Clifford benchmarks to application-motivated encoded Trotter circuits on hardware that supports in-circuit stabilizer checks, and ask whether these additional control capabilities yield a measurable reduction in logical error.

The main contributions of this paper are as follows:
\begin{itemize}
\item We construct GSE-encoded Trotter circuits compatible with CliNR and implement Shor-style stabilizer measurements that restrict fault propagation beyond a single code block.
\item We show experimentally that mid-circuit stabilizer measurement is essential to the observed improvement, yielding up to 54\% reduction in logical error rate relative to direct physical Trotter execution.
\item We demonstrate that machine-learning-guided stabilizer selection can identify effective verification operators and provide a practical route to optimizing the CliNR protocol.
\end{itemize}

The novelty of this work is twofold. First, we integrate GSE-based encoded Trotter circuits, CliNR, and Shor-style stabilizer verification into a single, experimentally validated framework for lower-error quantum simulation, and show that this combination reaches a regime in which CliNR provides a measurable advantage. Second, we provide an early application-driven demonstration of mid-circuit measurement on IonQ’s barium trapped-ion hardware, a Barium development system similar to the forthcoming IonQ Tempo line. We further show that the observed gain depends critically on timely in-circuit readout, rather than on additional verification circuitry alone, and that stabilizer selection can be optimized with machine learning rather than chosen heuristically. Taken together, these results outline a practical path toward lower-error encoded simulation on dynamic-circuit trapped-ion hardware.

The remainder of this paper is organized as follows. Section II reviews the relevant background on noise, error propagation, and Trotterized quantum simulation. Section III describes the noise-resilient circuit construction methods used in this work. Section IV presents the hardware platform, simulator framework, and calibrated noise model. Section V reports the experimental and simulation results, including the role of mid-circuit measurement and machine-learning-guided stabilizer selection. Section VI concludes.

\section{Background}

This section reviews the concepts needed for the remainder of the paper, with emphasis on quantum circuits and noise, product-formula Hamiltonian simulation, local fermion-to-qubit encodings, fault-tolerant stabilizer measurement, and Clifford noise reduction.

\subsection{Quantum circuits, noise, and error management}

Quantum computation is commonly described in the circuit model, in which qubits are initialized, transformed by one- and two-qubit gates, and finally measured~\cite{nielsen2010computation}. On present-day devices these operations are imperfect: coherent control error, stochastic gate error, decoherence, leakage, crosstalk, and readout error accumulate with circuit depth~\cite{nielsen2010computation,preskill2018NISQ}. Two-qubit gates are especially consequential because they usually dominate the infidelity budget and can propagate a single local fault into a correlated multiqubit error. This interplay between circuit depth and fault propagation is central to the present work.

The long-term remedy is quantum error correction (QEC), which encodes logical qubits into many physical qubits and repeatedly extracts syndrome information so that faults can be detected and corrected without destroying the logical state~\cite{campbell2017FTQC,svore2005localFTQC,gottesman1997thesis}. Practical QEC also requires decoding and architectural choices that balance qubit overhead, connectivity, and measurement cadence~\cite{bravyi2014MLD}. Recent experiments have begun to show logical performance at or beyond break-even~\cite{sivak2023realtime,google2025belowthreshold,bluvstein2026faulttolerant,paetznick2024logical}, but full fault tolerance remains expensive. This motivates lower-overhead strategies---including compilation, verification, and error-mitigation protocols---that improve accuracy before universal fault-tolerant operation is available~\cite{cai2023quantum}.

\subsection{Hamiltonian simulation and product-formula circuits}

A central task in quantum simulation is to approximate the real-time evolution
\begin{align}
U(t)=e^{-iHt}
\end{align}
generated by a Hamiltonian \(H\)~\cite{feynman1982simulating,lloyd1996universal,kassal2008polynomial,mcArdle2020quantum}. After mapping a fermionic Hamiltonian to qubits, one typically obtains
\begin{align}
H=\sum_{j=1}^{m} h_j P_j,
\end{align}
where each \(h_j \in \mathbb{R}\) and each \(P_j\) is a Pauli string~\cite{whitfield2011simulation,bravyi2002fermionic}. If the Pauli terms commute, the evolution factorizes exactly. In general they do not, and one instead uses product formulas such as the first-order Lie--Trotter approximation
\begin{align}
e^{-iHt}\approx \left(\prod_{j=1}^{m} e^{-i h_j P_j \Delta t}\right)^{N_T},
\qquad
\Delta t=t/N_T ,
\end{align}
where each repetition is a \emph{Trotter step} and \(N_T\) is the \emph{Trotter number}~\cite{suzuki1991general,childs2021theory}. Increasing \(N_T\) reduces product-formula error but also increases circuit depth, so algorithmic accuracy and hardware noise must be balanced against one another.

Product-formula methods remain attractive when low ancilla overhead and explicit circuit structure matter~\cite{wang2021simulation}. At the same time, the ordering and grouping of Pauli terms can substantially affect both Trotter error and gate count, especially for molecular Hamiltonians~\cite{tomesh2021optimized}. These considerations are directly relevant here because the circuits studied in this paper are encoded Trotter blocks whose error behavior depends on both the structure of the product formula and the way it is compiled to hardware.

\subsection{Encoded fermionic simulation with GSE and symplectic transvections}

The mapping from fermionic modes to qubits has a large impact on operator weight, locality, and error-detection opportunities. Canonical encodings such as Jordan--Wigner and Bravyi--Kitaev provide general fermion-to-qubit transformations but can yield high-weight or nonlocal Pauli operators for realistic systems~\cite{bravyi2002fermionic,whitfield2011simulation}. Local-encoding approaches seek to preserve more of the underlying interaction structure on the qubit register~\cite{verstraete2005mapping}. The GSE used in this work builds on this idea by associating fermionic structure with an interaction graph whose loop constraints become stabilizers~\cite{setia2019superfast,brown2025efficient}. In that sense, the encoding is not only a mapping of operators but also a source of native parity checks that can be exploited for error detection.

A separate challenge is how to implement logical evolutions on an encoded block without synthesizing each logical gate independently. Symplectic methods describe Clifford circuits and stabilizer-preserving transformations through binary symplectic matrices~\cite{rengaswamy2020logical}. Recent work connects this formalism directly to Trotter circuits via symplectic transvections, showing that logical product-formula blocks can be realized as physical circuits while preserving their ladder structure and stabilizer centralization~\cite{chen2025fault}. This connection is important for the present paper because it provides a route to encoded Trotter circuits that remain compatible with stabilizer-based verification.

\subsection{Fault-tolerant stabilizer measurement}

In stabilizer codes, commuting parity checks are measured to determine whether the state remains in the codespace and, if not, which error class has occurred~\cite{gottesman1997thesis,nielsen2010computation}. The measurement outcomes form the \emph{syndrome}. A key subtlety is that the syndrome-extraction circuit can itself introduce correlated faults, so stabilizer readout must be designed to limit error propagation to the data block~\cite{campbell2017FTQC,shor1996faulttolerant}. This is why fault-tolerant syndrome extraction is a central component of both QEC and verified noise-reduction protocols.

Shor's original scheme uses ancilla-assisted parity checks, typically with verified cat states, to ensure that a single ancilla fault does not spread into an uncorrectable correlated data error~\cite{shor1996faulttolerant}. Later work developed Steane-style extraction, constructions that interpolate between Shor and Steane ancillas, flag-qubit schemes that reduce ancilla overhead, and adaptive protocols that reduce the number of repeated rounds~\cite{huang2021syndromes,chao2018twoqubits,chao2020flag,tansuwannont2023adaptive}. In this paper, the phrase \emph{Shor-style stabilizer readout} refers broadly to ancilla-based parity extraction designed to confine fault propagation during stabilizer measurement.

\subsection{Noise reduction, error mitigation, and CliNR}

It is useful to distinguish full QEC from lower-overhead noise-reduction protocols. QEC repeatedly maintains the logical code space and, below threshold, supports arbitrarily long computations~\cite{campbell2017FTQC,svore2005localFTQC}. Error mitigation and noise reduction, by contrast, aim to improve the accuracy of a finite computation without guaranteeing scalable suppression of all errors. Examples include zero-noise extrapolation, probabilistic error cancellation, and symmetry-based postselection~\cite{temme2017error,li2017efficient,endo2018practical,mcArdle2019error,bonet2018low,cai2023quantum,takagi2022fundamental}. Historically, postselected and verified approaches were also studied as a way to extract useful computation from comparatively noisy components~\cite{knill2005noisy}.

CliNR belongs to this latter family but is specialized to Clifford subcircuits~\cite{delfosse2024lowcost,tham2025optimized}. The protocol prepares a Bell+Clifford resource state, measures a small set of its stabilizers, discards faulty preparations, and teleports the accepted Clifford operation onto the data register. Because the verification occurs off data, faults in the dominant Clifford block can be screened before they propagate through the rest of the algorithm. CliNR therefore sits between symmetry verification and full QEC: it does not protect an arbitrary computation in the threshold-theorem sense, but it can substantially reduce logical error for structured circuits at modest overhead. This is particularly attractive when the target algorithm contains substantial Clifford structure and when dynamic-circuit features such as mid-circuit measurement and reset are available to support repeated verification~\cite{delfosse2024lowcost,tham2025optimized,decross2023qubit,hothem2025measuring}.

\section{Noise Resilient Circuit Construction Methods}
This section describes how the circuits benchmarked in this work were assembled. We first synthesize an encoded Clifford Trotter block in the [[6,3,2]] Generalized Superfast Encoding (GSE), then convert that block into a CliNR resource-state circuit with ancilla-assisted stabilizer verification, and finally apply a graph-state recompilation tailored to the trapped-ion native entangling gate. Hardware calibration and the device noise model are deferred to Section~IV.

\subsection{Step I: Construct Trotter Circuits}
We target a single encoded Trotter block so that the possible outcomes are reduced and we can more easily detect advantages of different simulation constructions. Starting from the [[6,3,2]] GSE construction of Refs.~\cite{setia2019superfast,brown2025efficient} which is inherently confined to the even-occupation sector of three fermions given as
\[
\{(0,0,0),(1,1,0),(1,0,1),(0,1,1)\}.
\]
In the [[6,3,2]] GSE code, the logical occupation operators $B_i = (1- 2 a^{\dagger}_ia_i)$ are given as 
\[
B_0 = ZZIIII,\qquad B_1 = IIZZII,\qquad B_2 = IIIIZZ.
\]
and $A_{i,i+1}$ are defined in Ref.~\cite{brown2025efficient}.
The input state is prepared with a depth-3 circuit that initializes the logical occupation $(1,1,0)$ with all stabilizers $XYIIII, YXXYII, IIYXXY, IIIIYX$ having expectation value 1 as required.

The encoded block studied here implements the logical Pauli evolution
\begin{align}
\exp\left[i\frac{\pi}{4} B_0 A_{0,1} A_{1,2} B_2\right] = \exp\left[i \frac{\pi}{4} YZYYZY\right] .
\end{align}
where the $A_{i,j}$ operations are defined in Ref.~\cite{brown2025efficient}. Choosing the rotation angle $\pi/2$ makes the full block Clifford (as $R_Z(\pi/2)=S$), which allows the direct circuit, the CliNR resource state, and the verification circuits to be treated within the stabilizer formalism. The ideal output of the block is
\[
\frac{1}{\sqrt{2}}\left(\ket{110}+i\ket{011}\right).
\]

We synthesize the physical baseline with the CNOT ladder construction of Ref.~\cite{whitfield2011simulation}, while noting that the symplectic-transvection construction of Refs.~\cite{chen2025fault,rengaswamy2020logical} ensures the Trotter structure of the logical circuit at the physical level keeps the action inside the encoded code space. The resulting direct implementation, shown in Fig.~\ref{fig:physical_trotter}, is the reference circuit against which all verified variants are compared. The circuit description was then exported both to the hardware compilation flow and to the stabilizer-simulation flow used later in Section~IV-B through direct compilation with no further optimization applied to provide a fair comparison.

\subsection{Step II: Error-detected Trotter circuits}

Without using additional gadgets,  a fault inside the direct CNOT ladder can spread to a correlated data error that are possibly not detectable by the error detecting/correcting code.
Unlike previous work\cite{Chen_2025} that inserted additional flag gadgets directly into the data-block ladder, we verify the resource state associated with the encoded Trotter block before applying it.  Although we apply this design to a distance-2 GSE error detecting code, it is valid for any error correcting code and possibly provides a more scalable technique to apply general Clifford operations. The resource state is the CliNR protocol\cite{delfosse2024lowcost} Bell+Clifford with verification operators derived from the resource stabilizer state.

Concretely, for the $n=6$ encoded data block, the Bell+Clifford resource is a 12-qubit stabilizer state, and the experimentally tested verification schedules use one or two stabilizers drawn from that group. A shot is accepted only if every measured verification operator returns the $+1$ eigenvalue. Because the verification is performed before teleportation, preparation faults can be screened without coupling additional syndrome circuits directly to the encoded data register.

Each selected verification operator is measured with an ancilla-assisted parity-check circuit in the style of Shor syndrome extraction~\cite{shor1996faulttolerant,tansuwannont2023adaptive}. Alternative low-ancilla strategies based on flag qubits or compressed cat-state gadgets are possible~\cite{chao2018twoqubits,chao2020flag,prabhu2023syndrome}, but they would generally need to be co-designed with the specific error-correcting code. In the present methodology, the highest-risk entangling sequence is kept off the data block, and faulty resource preparations are rejected before they are teleported onto the encoded state.

\subsection{Step III: Use of CliNR with and without mid-circuit measurements}

We implement CliNR using the standard Bell+Clifford construction of Ref.~\cite{delfosse2024lowcost}. For the target Clifford block $C$, we prepare the corresponding Bell+Clifford resource state, measure a selected set of verification stabilizers, accept only shots with trivial verification syndrome, and then teleport $C$ onto the encoded data register. The Pauli feed-forward correction is
\begin{align}\label{eq.Q_correction}
Q=\prod_{i=1}^{n}\left(C^\dagger X_{2n+i}C\right)^{o_{n+i}}
\left(C^\dagger Z_{2n+i}C\right)^{o_i},
\end{align}
where $o_i$ and $o_{n+i}$ are the Bell-measurement outcomes. Figure~\ref{fig:CliNR_Circuit} shows the resulting circuit structure.

We construct two versions of this protocol. In the mid-circuit-measurement (MCM) version, the ancilla for each verification check is measured (along with a leakage check for the ancilla qubits) and reset immediately after the corresponding parity circuit, allowing the same ancilla pool to be reused across multiple verification rounds. In the delayed-readout version, the same verification entangling gates are applied but the ancilla measurements are postponed to the end of the circuit (with no leakage check). This isolates the timing of readout as the only methodological change while keeping the verification action itself fixed, and it mirrors the limitation of earlier trapped-ion CliNR demonstrations that lacked dynamic-circuit support~\cite{tham2025optimized}.

In both variants, failed verification leads to shot rejection and re-execution rather than to an attempt to repair the resource state in place. This keeps the protocol within the standard CliNR workflow while making the effect of true mid-circuit readout directly testable in hardware.

\subsection{Step IV: Integration of Error Detection and CliNR}
The integration strategy is to keep the data register encoded throughout and to execute the noisy Clifford block on a separate resource register. For an $n$-qubit data block, the Bell+Clifford resource occupies $2n$ qubits, and the addition of verification ancillas brings the compiled circuit width into the mid-$20$-qubit range for the instances studied here. Within this layout, the Bell measurement and correction stage act transversally across corresponding qubits, so accepted residual faults remain local on the data block to first order. By contrast, direct execution of the physical Trotter ladder can transform a single two-qubit fault into a correlated error that exceeds the detection capability of the [[6,3,2]] code.

This distinction is the methodological reason to combine GSE with CliNR in the present setting. In the direct circuit, the code only sees errors after they have already propagated through the ladder. In the verified circuit, the dominant Clifford subcircuit is first prepared and checked off-line, and only then teleported onto the encoded data. Within this encoded setting, CliNR need only suppress accepted resource-state errors below the detectability threshold of the code; it does not require full correction of every fault on the resource register.

For the hardware study, we therefore concentrated on shallow verification schedules: one mid-circuit stabilizer round, one delayed-readout stabilizer round, or one of each. This keeps the verification overhead within the gate-count and coherence budget of the device while still testing the central methodological question of the paper: whether the performance gain comes from off-data verification in general, or specifically from timely mid-circuit readout.

\subsection{Step V: Hardware-specific circuit compilation with graph resource states}

To better match the trapped-ion gate set, we recompile the Bell+Clifford resource state into graph-state realizations. We represent the target stabilizer resource by local-Clifford-equivalent graph states and navigate that family using local complementation~\cite{vandennest2004graphical}. This lets us replace a generic Clifford synthesis by a graph-state preparation dominated by CZ-type entangling operations plus single-qubit local Cliffords, which is a better match to the native ZZ interaction after basis changes.

In practice, we sample multiple local-Clifford-equivalent graph compilations of the target resource state, evaluate them with the same Clifford-simulation flow used later in Section~IV-B, and retain low-cost candidates for hardware execution. Candidate graph states are filtered to representatives with low CZ complexity~\cite{kumabe2025complexitygraphstatepreparationclifford}. 

\begin{table}[]
    \centering
    \begin{tabular}{|c|ccc|}
    \hline
        Clifford & \# Qubits & \# zz gates & \# gates \\
        \hline
        Physical Trotter & 6 & 16 & 173 \\
        CliNR Trotter & 26 & 58 & 580 \\
        CliNR Graph S1 & 26 & 52 & 475 \\  
        CliNR Graph S2 & 25 & 50 & 439 \\
        CliNR Graph S3 & 25 & 50 & 439 \\
        CliNR Graph S4 & 26 & 52 & 483 \\  
        CliNR Graph S5 & 26 & 58 & 559 \\
        CliNR Graph S6 & 25 & 54 & 479 \\
        \hline
    \end{tabular}
    \caption{Representative compiled widths and gate counts for the direct encoded Trotter circuit, the naive Bell+Clifford resource construction, and several local-Clifford-equivalent graph-state recompilations.}
    \label{tab:gates_qubits}
\end{table}

This graph-state representation also defines the stabilizer-search space used for verification. For each chosen graph compilation, candidate checks are drawn from the stabilizer group of the compiled resource state, and the same adjacency-matrix representation is later used in Section~V-C for machine-learning-guided stabilizer selection. In the future, this also makes it easier to try and use ML techniques to better prepare the resource state\cite{doherty2026faststabilizerstatepreparation}.

\section{Hardware Platform and Calibrated Simulation Framework}

The results in Section~V are supported by paired hardware and simulation studies. The hardware experiments establish the end-to-end behavior of the proposed circuits on a Barium development system similar to the forthcoming IonQ Tempo line, while the calibrated stabilizer simulations provide a device-level baseline for determining which observed trends are captured by the dominant modeled noise channels and which likely arise from additional hardware effects.

\subsection{Trapped-Ion Hardware Platform}

The experiments reported here were carried out on IonQ's trapped-ion quantum processors. We used a barium development system similar to the forthcoming IonQ Tempo line. The system supported up to 40 qubits and provided two features that are central to this work: mid-circuit measurement and highly reconfigurable, near--all-to-all two-qubit interactions mediated by collective motional modes. Ions were created via laser ablation and selective ionization and confined in compact, integrated vacuum packages using surface linear Paul traps. Universal control was implemented through two-photon Raman transitions driven by laser pulses, enabling arbitrary single-qubit rotations and entangling gates.

The primary systems demonstrated median direct randomized benchmarking fidelities for two-qubit gates of approximately 99.5\% and single-qubit gate fidelities of 99.99\%. Each mid-circuit measurement is paired with a leakage check. In our experiments, the leakage rate is 0.1\%.

For each circuit, the qubit-to-ion mapping was optimized over a subset of the 40-ion chain to maximize the average fidelity of the required two-qubit interactions. This has an important consequence for the comparison between direct Trotter execution and CliNR. The bare Trotter circuits use only six qubits and can therefore be placed on the best-performing entangling pairs in the chain. The CliNR circuits are substantially wider and must occupy a larger subset of ions, which in turn requires the use of some two-qubit pairs whose benchmarked fidelities are lower than those available to the six-qubit baseline. The hardware comparison is therefore conservative with respect to CliNR: any improvement observed for the verified circuits is obtained despite a less favorable qubit-mapping environment.

To compile the circuits to the native hardware gate set, we converted each single- and two-qubit gate into the corresponding sequence of native GPI, GPI2, and ZZ gates without further optimization. The same native-gate description is used in the simulation flow described next, allowing the calibrated model to be compared directly against hardware.

\subsection{Simulation Framework and Calibrated Noise Model} \label{sec:noise_model}

Because all circuits in this study are Clifford, they can be simulated efficiently via the Gottesman--Knill theorem. We used NVIDIA's CUDA-Q cuQuantum cuStabilizer~\cite{cuda_q,NVIDIA2026cuStabilizer}, a GPU-accelerated stabilizer simulator, with circuits described in the \emph{Stim} format~\cite{gidney2021stim}. cuStabilizer is highly optimized for NVIDIA GPUs through bit‑packed stabilizer representations and batched GF(2) operations, enabling high throughput for large numbers of shots. Stabilizer simulation scales polynomially in qubit count, allowing $8.4 \times 10^{4}$ shots per circuit for the $25$--$34$ qubit width sweeps and up to $5 \times 10^{5}$ shots for the higher-precision $25$--$26$ qubit studies.

Circuits in IonQ's native gate format ($GPI$, $GPI2$, $ZZ$) were translated to Stim. Virtual $Z$ rotations ($GZ$) were absorbed into the phases of subsequent native gates and contribute no noise, matching their virtual implementation on hardware. Mid-circuit measurements were followed by a reset to $\ket{0}$, and outcomes were recorded for post-selection on stabilizer syndromes.

The noise model \texttt{tempo-1} was calibrated to IonQ's 
Barium development system similar to the forthcoming IonQ Tempo line and consists of four Pauli channels; rates are quoted in parts per ten thousand ($1~\text{pptt} = 10^{-4}$).
\begin{itemize}
  \item \textbf{Single-qubit gate error.} A pure dephasing channel applies a $Z$ error with probability $p_Z = 2.55 \times 10^{-4}$ after each single-qubit gate, reflecting the dominance of Raman-induced phase noise over bit-flip errors on this platform.
  \item \textbf{Two-qubit gate error.} After each $ZZ$ gate, an independent $Z$ error is applied to each qubit with probability equal to half the direct randomized benchmarking (DRB) rate of the corresponding ion pair, calibrated across 780 pairs of a 40-ion chain. Because the qubit-to-ion mapping is optimized per circuit width, the six-qubit bare Trotter circuit lands on the cleanest pairs (usage-weighted mean $\sim\!41$~pptt), while wider CliNR circuits span more ions and necessarily include noisier pairs (mean $\sim\!59$~pptt).
  \item \textbf{Idle dephasing.} Any qubit idle during a circuit layer accumulates a $Z$ error with probability
  \[
    p_z = \frac{1 - e^{-\Delta t / T_2^*}}{2},
  \]
  where $T_2^* = 1.5~\text{s}$ and $\Delta t$ is the layer duration: $130~\mu\text{s}$ (single-qubit), $950~\mu\text{s}$ (two-qubit), or $400~\mu\text{s}$ (measurement).
  \item \textbf{Readout error.} Each measurement outcome is independently flipped with probability $p_\text{spam} = 1.2 \times 10^{-3}$, scaled from the published single-ion Ba-133 readout fidelity~\cite{IonQ2022barium}.
\end{itemize}
Per-pair rates capture the qubit-mapping asymmetry between narrow and wide circuits described in Section~IV-A. The model omits crosstalk and non-Markovian effects, so any features visible in hardware but absent from simulation point to mechanisms beyond this Pauli-channel description.

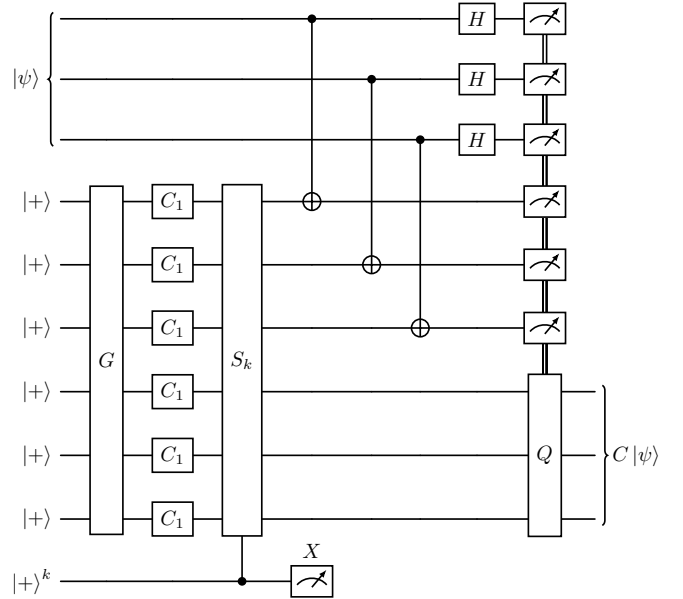
\begin{figure}[htbp]
\resizebox{1\linewidth}{!}{%
\begin{quantikz}
\lstick[3]{$\ket{\psi}$} & & & & \ctrl{3}& & & \gate{H} & \meter{}\wire[d][6]{c} \\
& & & & &\ctrl{3} & & \gate{H} & \meter{}\wire[d][5]{c}\\
& & & & & & \ctrl{3}& \gate{H} & \meter{}\wire[d][4]{c}\\
\lstick{$\ket{+}$}& \gate[6]{G} & \gate{C_1} & \gate[6]{S_k} & \targ{} & & & & \meter{}\wire[d][3]{c}  \\
\lstick{$\ket{+}$}& & \gate{C_1} & &  &\targ{} & & & \meter{}\wire[d][2]{c}\\
\lstick{$\ket{+}$}& & \gate{C_1} & & &  & \targ{}& & \meter{}\wire[d][1]{c}\\
\lstick{$\ket{+}$}& & \gate{C_1} & & & & & & \gate[3]{Q} & \rstick[3]{$C\ket{\psi}$}\\
\lstick{$\ket{+}$}& & \gate{C_1} & & & & & & &\\
\lstick{$\ket{+}$}& & \gate{C_1} & & & & & & & \\
\lstick{$\ket{+}^k$}&  & & \ctrl{-1} & \meter{X}& \wireoverride{n}
\end{quantikz}
}

\caption{Illustrative CliNR circuit with graph state resource (G) and single qubit clifford ($C_1$) preparation for a three qubit system $\ket{\psi}$ along with  one mid-circuit measured stabilizer $S_k$ of length $k$.  Unlike the original implementation of CliNR, the stabilizer measurement is performed using Shor-style measurements with $\ket{+}^k$ representing a fault-tolerantly prepared cat state of length $k$. The feedforward correction $Q$, defined in Eq.\ref{eq.Q_correction}, is applied in post-processing in this work.\label{fig:CliNR_Circuit}}
\end{figure}

\begin{figure}[htbp]
    \centering
\resizebox{1\linewidth}{!}{%
    \begin{quantikz}
\lstick{$\ket{0}$} & \gate{R_X(\theta)} & \ctrl{1} &  &  &  &  &  &  &  &  &  & \ctrl{1} & \gate{R_X(-\theta)} \\
\lstick{$\ket{0}$} &  & \targ{} & \ctrl{1} &  &  &  &  &  &  &  & \ctrl{1} & \targ{} &  \\
\lstick{$\ket{0}$} & \gate{R_X(\theta)} &  & \targ{} & \ctrl{1} &  &  &  &  &  & \ctrl{1} & \targ{} &  & \gate{R_X(-\theta)} \\
\lstick{$\ket{0}$} & \gate{R_X(\theta)} &  &  & \targ{} & \ctrl{1} &  &  &  & \ctrl{1} & \targ{} &  &  & \gate{R_X(-\theta)} \\
\lstick{$\ket{0}$} &  &  &  &  & \targ{} & \ctrl{1} &  & \ctrl{1} & \targ{} &  &  &  &  \\
\lstick{$\ket{0}$} & \gate{R_X(\theta)} &  &  &  &  & \targ{} & \gate{R_Z(\theta)} & \targ{} &  &  &  &  & \gate{R_X(-\theta)}
\end{quantikz}
}
    \caption{The Physical Trotter (C) implementation of the logical operation $\sigma_{1}\sigma_{5}=YZYYZY$ where $\theta=\pi/2$. In general, single-qubit faults occurring between CNOT gates can propagate to high-weight faults when implementing this circuit.}
    \label{fig:physical_trotter}
\end{figure}
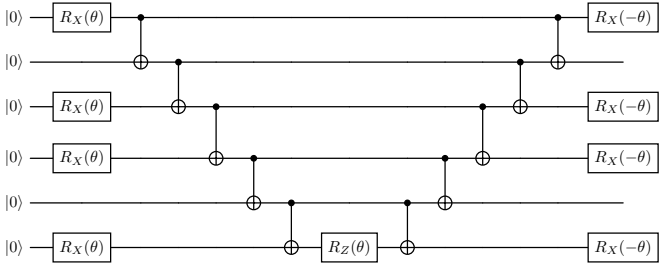

\section{Evaluation of Noise-Reduction Performance}

This section evaluates our noise-reduction techniques on hardware and in simulation. We first compare various circuit executions on the Barium development system, then use calibrated simulations to interpret the observed trends, and finally demonstrate how to use machine learning to select the best stabilizers.

\subsection{Hardware Results}
We compare the error rates of the physical Trotter circuit with the CliNR circuit implementing the same encoded dynamics on trapped ion hardware. The CliNR and bare Trotter implementations are shown in Fig.~\ref{fig:CliNR_Circuit} and Fig.~\ref{fig:physical_trotter}, respectively. Our CliNR implementation uses one stabilizer measured mid-circuit and a second stabilizer whose ancilla readout is deferred to the end of the circuit.

\begin{figure}[htbp]
\centering\includegraphics[width=0.90\linewidth]{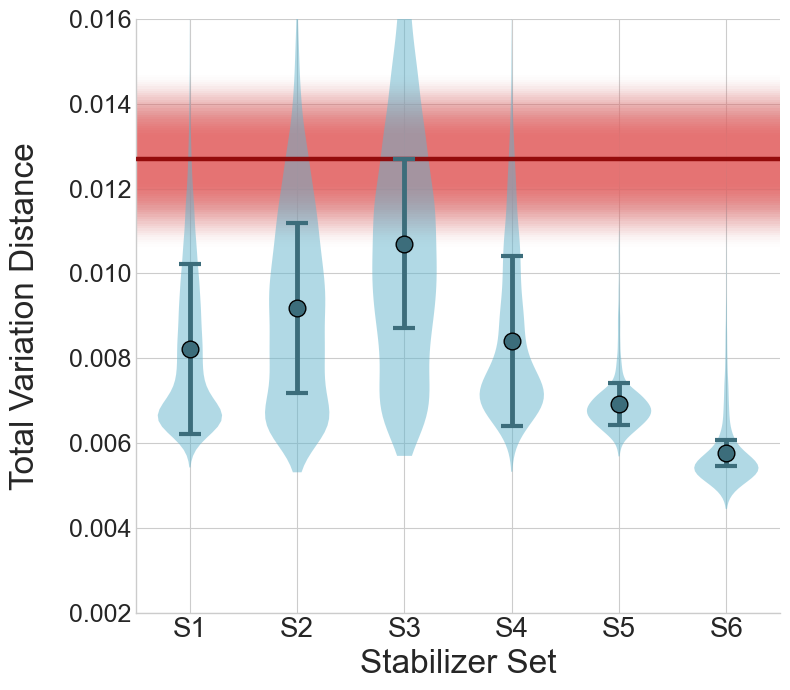}
\caption{Hardware results for the stabilizer pairs listed in Table~\ref{tab:placeholder} for the combined Bell+Clifford state to perform the Trotter evolution. Each pair improves on the physical Trotter baseline (red line with std error), although the magnitude of the improvement depends on the chosen stabilizers.}
\label{fig:exp_random}
\end{figure}

To quantify error rates, we use the total variation distance (TVD) between the measured distribution and the ideal distribution. For this experiment, the TVD can be decomposed into two contributions: (1) fraction of states with incorrect occupation, and (2) bias between the (1,1,0) and (0,1,1) outcomes. We evaluated six randomly chosen stabilizer pairs. The results, shown in Fig.~\ref{fig:exp_random}, indicate that every stabilizer choice yields lower error than the physical Trotter circuit. In particular, stabilizers S5 and S6 perform significantly better than the physical Trotter circuit. The best result is obtained with stabilizer S6, where the TVD is reduced from $0.0125$ to $0.0059$, a relative improvement of $54\%$.

Table~\ref{tab:bias_} breaks the TVD into its two components for each of the six stabilizer pairs. In every case, the CliNR circuit reduces both the bias within the correct subspace and the probability of measuring a state in the wrong subspace relative to the physical Trotter circuit. The incorrect-state rate drops by about $30\%$ for all stabilizer choices, and by up to $42\%$ in the best case. The bias between the $(1,1,0)$ and $(0,1,1)$ outcomes varies more strongly across circuits. We quantify this by computing the probability that an observed imbalance, $P(1,1,0) \neq P(0,1,1)$, cannot be attributed to statistical fluctuations alone. While some CliNR choices still exhibit noticeable bias, this probability is lower in every case than for the physical Trotter circuit. For the physical Trotter circuit, the probability that the observed imbalance occurs without an underlying bias is only $10^{-5}$, whereas the best CliNR circuit shows much weaker evidence of such a bias. This bias corresponds to a non-Clifford state and is therefore not captured by standard Clifford noise models.

\begin{table}[]
    \centering
    \begin{tabular}{|c|cc|}
    \hline
         & $P\left((1,1,0) \ne (0,1,1)\right)$  & Incorrect State  \\
         \hline
      Physical Trotter   & 99.999\% & 0.91\% \\
      S1 & 95.80\% &  0.65\%  \\
      S2 & 97.85\% &  0.64\% \\
      S3 & 99.59\% &  0.66\% \\
      S4 & 86.65\% &  0.70\% \\
      S5 & 33.81\% & 0.67\%\\
      S6 & 25.30\% & 0.53\% \\
      \hline
    \end{tabular}
    \caption{A breakdown of the TVD errors for each circuit for the decoded logical states. The first contribution comes from the biasing of the measurements of the (1,1,0) and (0,1,1) states. The physical Trotter circuit is by far the most biased. The second contribution is due to the appearance of incorrect states (1,0,1) and (0,0,0). All stabilizers also reduce the amount of incorrect states by 25\% to 40\%.}
    \label{tab:bias_}
\end{table}

\begin{table}[]
    \centering
    \begin{tabular}{|c|cc|}
    \hline
    Label & Stabilizer 1 & Stabilizer 2 \\
    \hline
    S1 & $X_6 Y_8 Z_{12} Z_{13} Y_{15} Z_{16} Y_{17}$ & $Z_6 X_9 Z_{12} X_{15}$ \\
    S2 & $Y_{10} Z_{11} Y_{16} Z_{17}$ & $Z_6 Z_8 Z_{10} Z_{12} Z_{14} Z_{16}$ \\
    S3 & $Y_{10} Z_{11} Y_{16} Z_{17}$ & $Z_6 Z_7 Z_{11} Z_{12} Z_{13} Z_{17}$ \\
    S4 & $Z_8 Y_{10} Z_{14} Y_{16}$ & $Y_6 X_8 Y_9 Z_{13} Z_{14} Z_{16} Y_{17}$ \\
    S5 & $Y_6 Y_8 Y_9 Y_{10} Z_{13} X_{16} Y_{17}$ & $X_{6} Y_{9} Z_{12} Z_{13} Y_{14} Z_{16} Y_{17}$ \\
    S6 & $Z_6 X_7 Z_{10} Z_{12} X_{13} Z_{16}$ & $Z_6 Z_8 Y_{11} Z_{12} Z_{14} Y_{17}$  \\
    \hline
    \end{tabular}
    \caption{The stabilizers used for the CliNR state checks. Stabilizer 1 was measured mid-circuit, while Stabilizer 2 had the gates applied mid-circuit but the ancilla measurement deferred to the end.}
    \label{tab:placeholder}
\end{table}

\begin{figure}[htbp]
\centering\includegraphics[width=\linewidth]{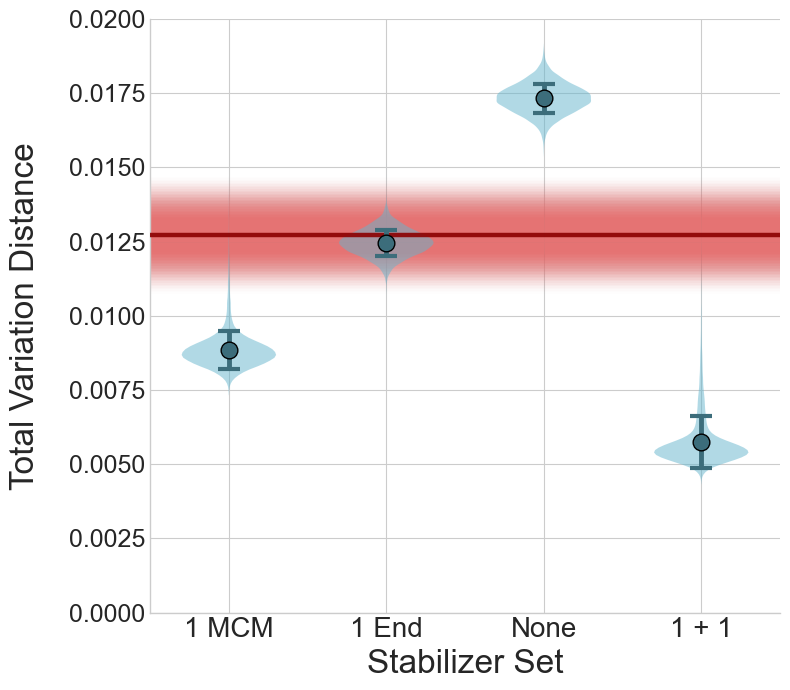}
\caption{Comparison of one stabilizer measured mid-circuit (1 MCM), one stabilizer with ancilla readout deferred to the end (1 End), no stabilizer measurements (None), and one mid-circuit plus one end-of-circuit stabilizer (1+1). Mid-circuit readout is required to obtain the strongest improvement. As expected, not performing stabilizer measurements performs worst, while adding a second stabilizer can further improve the result..}
\label{fig:mid-v-end}
\end{figure}

Our results also highlight the importance of mid-circuit stabilizer measurements. In Fig.~\ref{fig:mid-v-end}, we compare a single-stabilizer CliNR circuit using the stabilizer $X_6 Y_8 Z_{12} Z_{13} Y_{15} Z_{16} Y_{17}$ under two readout schedules: mid-circuit measurement (MCM) and end-of-circuit measurement (ECM). We also include a CliNR circuit with no stabilizer measurement but with the same implementation overhead. 

Both ECM and MCM outperform the no-stabilizer baseline, confirming the benefits of error detection. More importantly, the MCM circuit outperforms ECM, and only the MCM case yields a statistically significant improvement over the physical Trotter circuit for a single stabilizer round. This behavior contrasts with earlier CliNR experiments in Refs.~\cite{delfosse2024lowcost,tham2025optimized}, where end-of-circuit stabilizer measurements produced approximately breakeven performance in deep random circuits. As a further point, ignoring the MCM leakage check does not significantly change the result, with the error only increasing from $0.884$ to $0.898$ TVD. Figure~\ref{fig:mid-v-end} also includes the best two-stabilizer result from Fig.~\ref{fig:exp_random} (S6), illustrating that multiple stabilizer checks can further improve performance. 

Overall, these hardware results show that the barium-based trapped-ion system used here operates below the breakeven point for our CliNR+GSE-based Trotter circuits, and that the system's mid-circuit measurement capability is critical to realizing this improvement.

\subsection{Simulation Results}

We use cuQuantum simulations to help interpret the error processes observed in the experiment. The noise model described in Sec.~\ref{sec:noise_model} qualitatively reproduces most of the experimental trends, although some features likely require a more detailed noise model.

\begin{figure}[htbp]
\centering\includegraphics[width=0.9\linewidth]{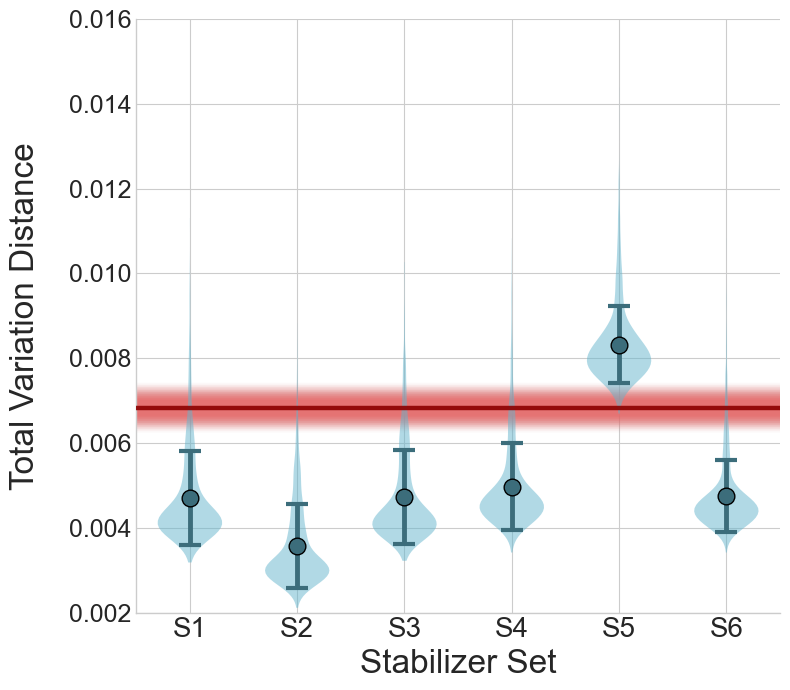}
\caption{Noisy-simulation results for the same random stabilizer pairs as in Fig.~\ref{fig:exp_random}. The calibrated model captures the overall improvement trend, but not every stabilizer-dependent detail.}
\label{fig_label1z}
\end{figure}

Figure~\ref{fig_label1z} shows noisy simulations for the same randomly selected stabilizer pairs used in the hardware experiments. Nearly all pairs lie below the breakeven error rate, with S5 as the only clear outlier and the main point of disagreement with experiment. In both the CliNR and physical Trotter simulations, the absolute error rates are approximately $30\%$ lower than in hardware, suggesting the presence of additional noise sources not captured by our model.

\begin{figure}[htbp]
\centering\includegraphics[width=\linewidth]{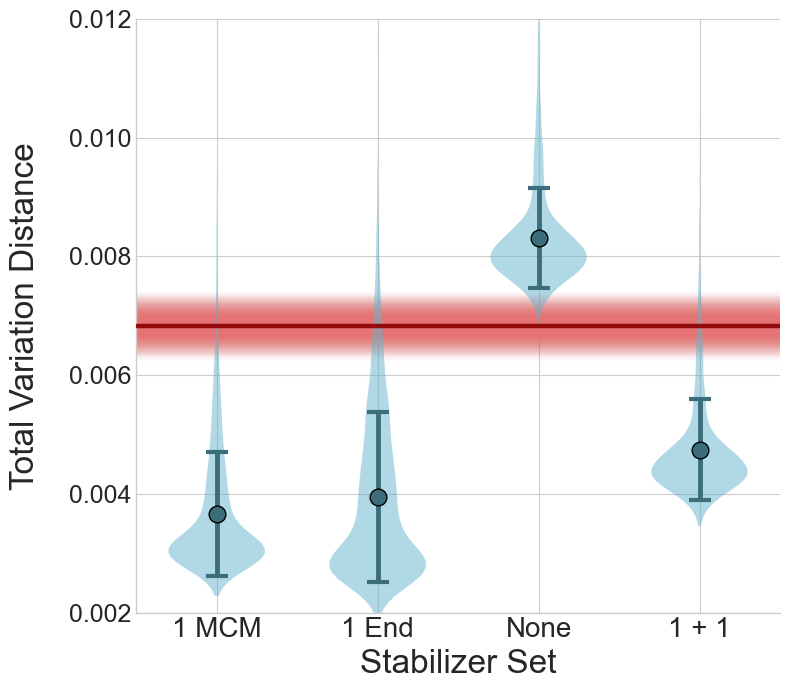}
\caption{Simulation of the circuits in Fig.~\ref{fig:mid-v-end}. Although the simulations capture the benefit of stabilizer measurements (1MCM, 1 End and 1+1) compared to simulations without stabilizers (None), we do not observe the additional benefit of performing mid-circuit measurements. This is likely due to the standard circuit-level noise model missing error mechanisms such as crosstalk.}
\label{fig:sim_mcm_v_end}
\end{figure}

We also simulated the circuits from Fig.~\ref{fig:mid-v-end}, where the hardware experiments demonstrated the importance of mid-circuit measurements. The noisy simulation results, shown in Fig.~\ref{fig:sim_mcm_v_end}, exhibit the expected improvement from applying stabilizer measurements relative to both the physical Trotter circuit and the CliNR circuit without stabilizer measurements. However, unlike the hardware data, the simulations do not show any additional benefit when the stabilizer is measured mid-circuit rather than at the end of the circuit.

We attribute this difference to the simplicity of the idle-error model. The only errors applied to the ancilla qubits are $T_2$-type idle errors modeled as $Z$ gates, and these commute with the measurement operations. As a result, measurement timing has no effect in the simulation. The hardware sensitivity to delayed readout therefore points to additional noise processes not captured by the model, such as crosstalk or other non-Pauli effects. One plausible interpretation is that mid-circuit measurement allows faults arising within a CliNR block to be detected before they affect later portions of the circuit.

\begin{figure}[htbp]
\centering\includegraphics[width=\linewidth]{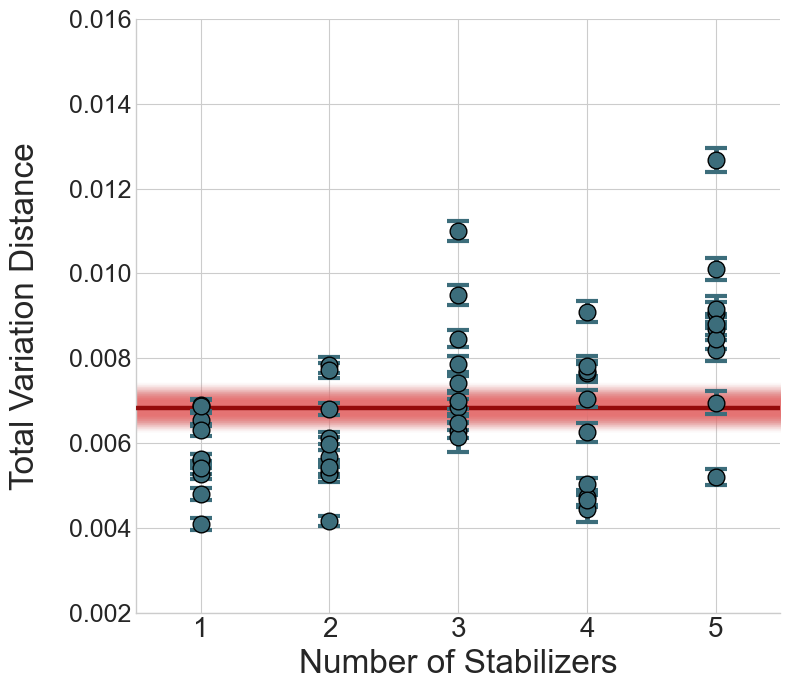}
\caption{Simulation results for different numbers of mid-circuit stabilizer measurement rounds using randomly selected stabilizers with weight $<8$. A larger number of stabilizers can produce lower error, but this is not guaranteed.}
\label{fig:sim_1_to_4}
\end{figure}

Finally, we performed noisy simulations for a broad family of circuits with between one and five stabilizers measured in series. The results, shown in Fig.~\ref{fig:sim_1_to_4}, emphasize that good performance depends not only on the number of stabilizers measured, but also on which stabilizers are chosen.

Even at fixed stabilizer count, the TVD varies substantially across circuits. Most stabilizer sets outperform the physical Trotter circuit, but some poorly chosen sets do not. The simulations suggest that the system is near the point where additional verification overhead begins to offset the benefit of extra error detection. By contrast, the hardware results suggest that the experiment may still be below that point, since measuring more stabilizers can further reduce the error rate as shown in Fig.~\ref{fig:mid-v-end}. Overall, these results show that careful stabilizer selection is essential for achieving improved performance.

\subsection{Machine-Learning-Guided Stabilizer Selection}

The six stabilizer pairs evaluated in Section~V-A were drawn at random from a very large candidate space. While most choices improve on the physical Trotter implementation, our experimental results demonstrate that the verification sequence---both the number of checks $r$ and the specific stabilizers chosen---has a direct impact on the noise reduction achieved by CliNR. For the 12-qubit resource state studied here (for  $n=6$ data qubits) with $r=2$ stabilizer measurements, the candidate set for a single graph compilation already contains $\sim8 \times 10^{6}$ stabilizer pairs. Brute-force screening with Clifford-circuit simulation becomes prohibitive as $n$, $r$, or the number of equivalent graph-state compilations grow. The tabu search of~\cite{tham2025optimized} prunes this space heuristically but still simulates every candidate it visits. A surrogate model that predicts stabilizer quality without simulation is therefore complementary to tabu search. It can serve as a standalone fast filter over the candidate space, or as an inner cost function inside a tabu-style outer loop.

\begin{figure}[htbp]
  \centering
  \includegraphics[width=0.9\linewidth]{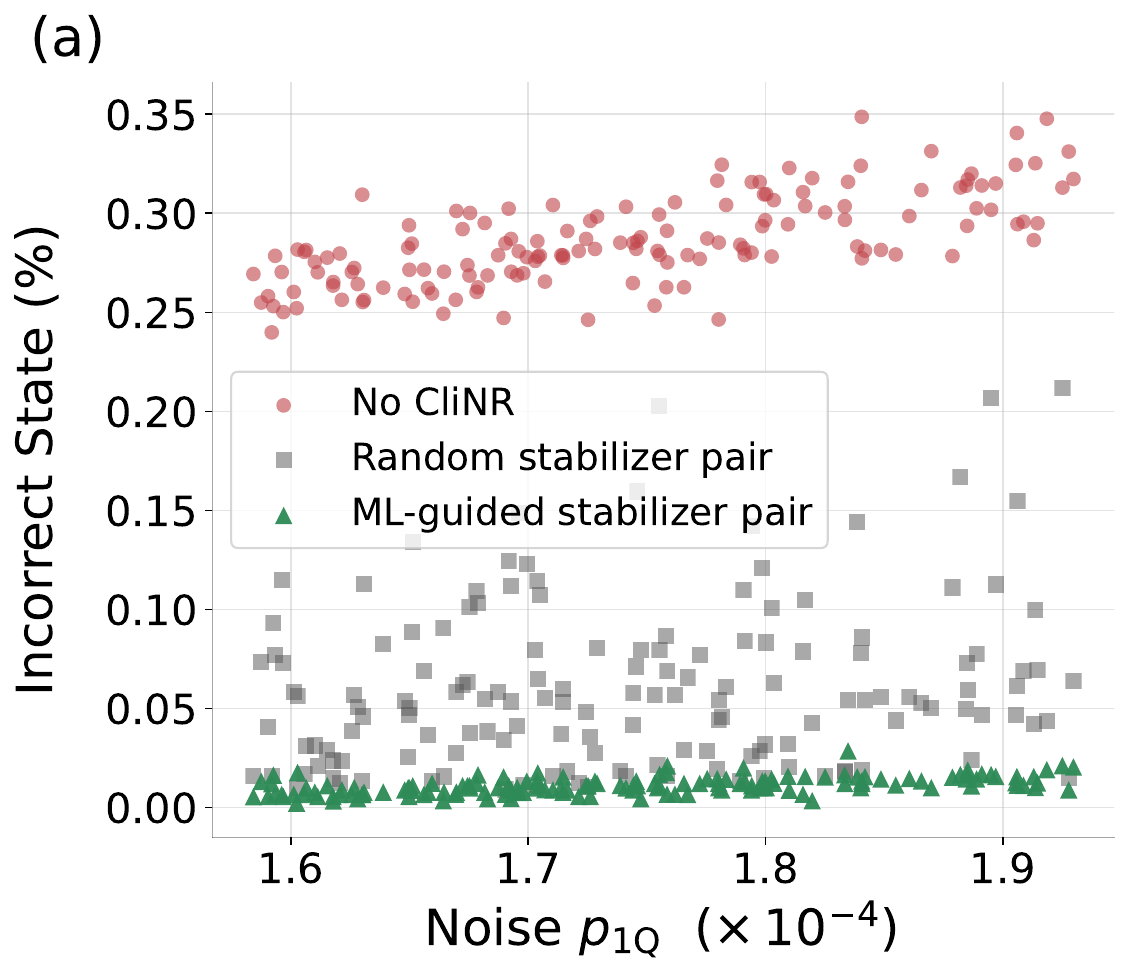}
  \includegraphics[width=0.9\linewidth]{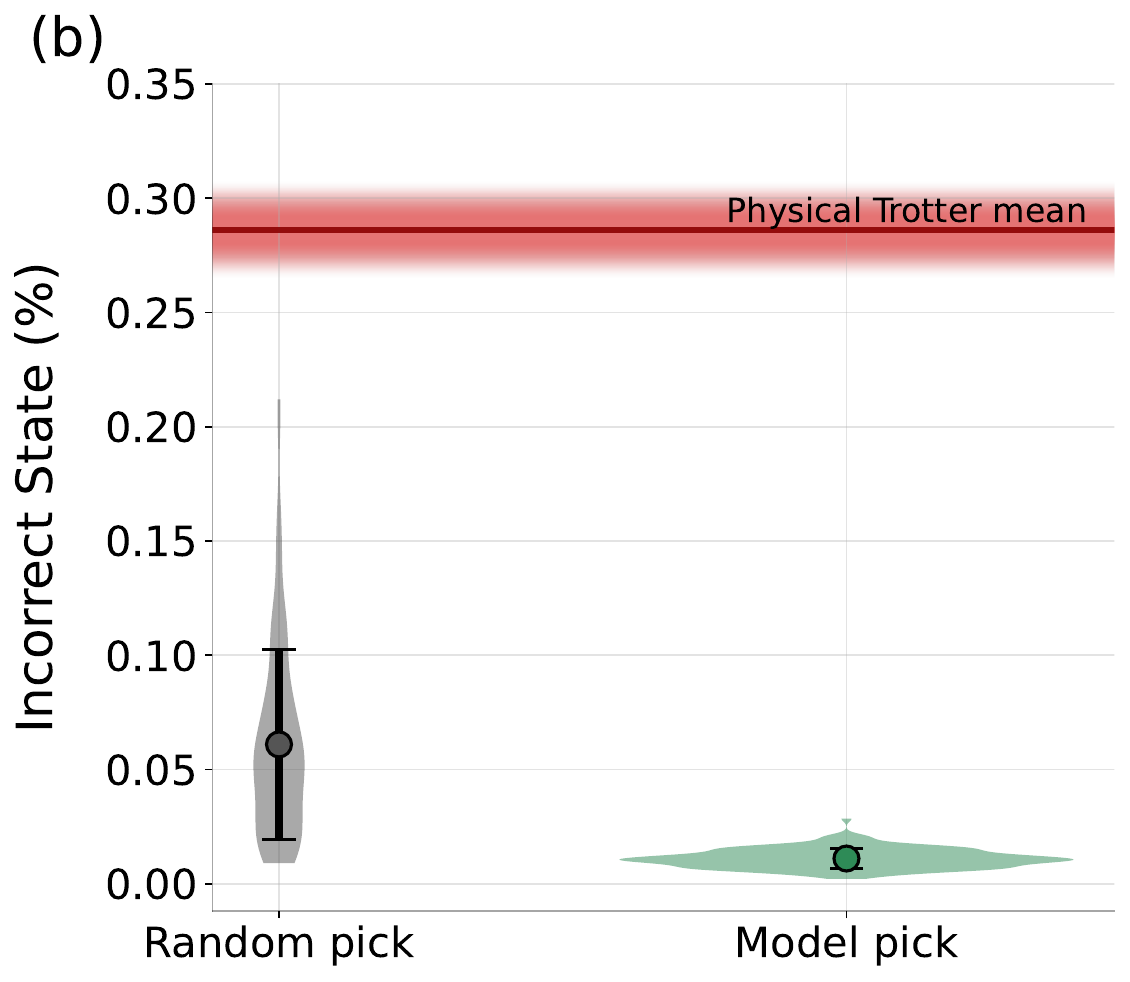}
  \caption{(a) Incorrect-state probability, $P_{\text{fail}}(\%)=(P(0,0,0) + P(1,0,1))\times 100$, across 150 independent simulation trials as a function of noise level $p_{1q}$, comparing physical Trotter circuit implementation (no CliNR), CliNR with a randomly drawn stabilizer pair, and CliNR with a model-guided stabilizer pair choice. Each point represents one independent trial ($10^5$ shots). (b) Distribution of incorrect-state probability across the same 150 trials for the two CliNR selection strategies. The red line shows the physical Trotter baseline. The model achieves a lower failure rate in 149/150 trials (99.3\%), with a mean reduction of 72.5\% relative to random selection.}
  \label{fig:ml_winrate}
\end{figure}

As a first step in this direction, we trained a Graph Attention Network (GAT)~\cite{velickovic2018graph} to predict the failure probability for a given verification sequence on a given graph state compilation of the Bell+Clifford CliNR resource state. Here, failure probability refers to the same metric as the ``Incorrect State'' column of Table~\ref{tab:bias_}, $P_{\text{fail}}=P(0,0,0) + P(1,0,1)$. The model takes as input the resource-state adjacency matrix $A \in \{0,1\}^{12 \times 12}$, the full Pauli strings of both selected stabilizers (each encoded as a $24$-dimensional binary vector specifying the $X/Y/Z$ type at every qubit), and the noise level.

A graph neural network operating directly on the adjacency matrix is a natural architectural choice for this problem. We find that representing each stabilizer by its full Pauli string is essential: a support mask that encodes only which qubits a stabilizer acts on achieves Spearman $\rho = 0.55$ on held-out data, versus $\rho = 0.74$ with full Pauli strings. Two stabilizers with identical support but different Pauli types detect different error patterns, and the model learns this distinction. An MLP baseline on the same flattened inputs ($\rho = 0.57$) and a graph convolutional network with fixed aggregation weights ($\rho = 0.71$) both underperform, indicating that learned attention over graph neighbors contributes to the model's performance.

We generated the training data with a simplified symmetric depolarizing noise model ($p_{2q} = 10\,p_{1q}$, $p_{\text{meas}} = p_{1q}$, no $Z$-bias, no idle errors), distinct from the calibrated tempo-1 model of Section~IV-B. The ML results should therefore be interpreted as a proof of concept under simplified noise; retraining under calibrated device noise is the immediate next step. The dataset comprises 57{,}536 samples drawn from 58 distinct CZ-minimal graph compilations of the resource state, with 992 unique stabilizer pairs per graph and $10^{5}$ shots per sample. We sampled the noise parameter from a $\pm 10\%$ band around a fixed operating point to prevent noise level variation from dominating the stabilizer-quality signal during training.

We trained the model on an NVIDIA GH200 GPU and evaluated it with a by-graph split to test generalization to new graph compilations. One graph compilation is held out entirely from training, and the model is scored on all 992 stabilizer pairs on the unseen graph. The model achieves Spearman $\rho = 0.74$ on the held-out graph. To further assess operational utility, we ran 150 independent trials. Fig.~\ref{fig:ml_winrate} shows our results. In each trial, a fresh graph compilation is generated, the model scores $10^{5}$ randomly drawn candidate pairs without running costly stabilizer simulation, the highest-scoring pair is simulated once, and the result is compared against a randomly drawn baseline pair simulated under the same shot budget. The model-guided selection achieves a lower failure rate in 149/150 trials (99.3\%) with a mean improvement of $72.5\%$ over random.

We have not tested generalization to different Clifford circuits ($C$), system sizes ($n$), or number of stabilizer measurements ($r$) in this work. A classifier variant trained directly to identify high-quality stabilizer pairs, retraining under calibrated noise matched to the hardware model, and coupling the model with tabu search as an inner cost function are natural directions for future work.

\section{Conclusion}
We presented a device-matched framework for reducing errors in encoded Trotter circuits for quantum simulation on trapped-ion hardware. By combining symplectic-transvection-based Trotter synthesis in the [[6,3,2]] Generalized Superfast Encoding with Clifford Noise Reduction and Shor-style stabilizer verification, we obtain a clear improvement over direct physical Trotter execution. The main experimental punchline is that mid-circuit stabilizer measurement is essential: with in-circuit verification, the logical error is reduced by up to 54\%, whereas delaying measurement to the end removes most of the benefit. These results show that encoding-native verification can deliver meaningful accuracy gains for application-motivated quantum simulation without the full overhead of quantum error correction.

A second conclusion is that both device details and verification choices matter. Our calibrated simulations reproduce much of the stabilizer-dependent behavior, but they do not fully capture the hardware advantage of mid-circuit readout, indicating that additional effects remain relevant in this regime. We also find that stabilizer choice is an important optimization lever, and our proof-of-concept machine-learning study shows that stronger verification sequences can be identified systematically rather than chosen heuristically. Taken together, these results move CliNR beyond generic Clifford benchmarks and show that dynamic-circuit capabilities can translate native stabilizer structure into measurable gains on trapped-ion hardware.

Future work will extend this framework to non-Clifford Trotter circuits, where CliNR verifies the Clifford portions surrounding a physical rotation gate in the generalized symplectic-transvection construction \cite{chen2025fault}. An equally important next step is tighter integration of accept/reject resource-state preparation with real-time feed-forward so that post-selection overhead is reduced during execution. In particular, coupling this workflow through NVQLink \cite{caldwell2025platformarchitecturetightcoupling} is a promising systems path for retaining the accuracy gains demonstrated here while making the overall methodology substantially more practical.

\section*{Acknowledgment}
We thank Mike Goldman, Ken Wright, and Neal Pisenti for their assistance in running CliNR circuits on the Barium development system and for their help in mapping our circuits to the best-performing qubit pairs. We also thank Guen Prawiroatmodjo and Amrit Poudel for assisting with the construction of realistic noise models, and Nicolas Delfosse and Edwin Tham for fruitful discussions regarding noise mitigation techniques. During preparation of all sections of this manuscript the authors used OpenAI ChatGPT Pro 5.4 for grammar enhancement, readability-focused editing, and literature-review support; all technical content, interpretations, citations, and final editorial decisions were made and verified by the authors.

\bibliographystyle{IEEEtran}
\bibliography{references}

\end{document}